\documentclass[aps,prl,reprint,nofootinbib,longbibliography,floatfix]{revtex4-2}
\usepackage[T1]{fontenc}
\usepackage[utf8]{inputenc}
\usepackage{amsmath,amssymb,amsthm,mathtools,graphicx,txfonts,microtype}
\usepackage{capt-of}
\usepackage[hidelinks]{hyperref}
\newcommand{\bra}[1]{\langle #1\rvert}
\newcommand{\lettertitle}{Superadditivity of Entanglement Distillation with Limited Forward Classical Communication}
\hypersetup{pdftitle={\lettertitle},pdfauthor={Anthony M. Polloreno}}
\newcommand{\ket}[1]{\lvert #1\rangle}
\newcommand{\Tr}{\operatorname{Tr}}
\newcommand{\diag}{\operatorname{diag}}
\newcommand{\fp}{\mathrm{fp}}
\newtheoremstyle{qit}{\topsep}{\topsep}{\normalfont}{\parindent}
  {\itshape}{.---}{0pt}{}
\theoremstyle{qit}
\newtheorem{theorem}{Theorem}
\begin{document}
\setlength{\textfloatsep}{10pt plus 2pt minus 2pt}
\setlength{\dbltextfloatsep}{10pt plus 2pt minus 2pt}
\setcounter{secnumdepth}{-1}
\title{\lettertitle}
\author{Anthony M. Polloreno}
\email{ampolloreno@gmail.com}
\affiliation{Independent Researcher}
\date{November 29, 2026}

\begin{abstract}
We present a degradable state for which pairing copies improves Bell-pair yield with limited forward bits, despite separate processing reaching the same maximum at the same minimum cost. Part of our protocol uses a state with zero one-way distillable entanglement, which supplies a memory whose noise the receiver partly knows. We use Bloch-sphere geometry to argue that erasing a coin flip record can hide more about a signal's label than it hides about the signal itself, yielding a mixed quantum memory which outscores every flagged pure preparation in any finite dimension. We generalize our argument to prove that every collective gain on independent inputs yields a one-input test in which erasing a classical key raises the information bottleneck score.
\end{abstract}
\maketitle
\setlength{\abovedisplayskip}{6pt plus 2pt minus 3pt}
\setlength{\belowdisplayskip}{6pt plus 2pt minus 3pt}
\setlength{\abovedisplayshortskip}{6pt plus 2pt minus 3pt}
\setlength{\belowdisplayshortskip}{6pt plus 2pt minus 3pt}

If separate processing attains the maximum yield of Bell pairs with free forward communication, can joint processing still help when Alice can send only $c$ bits per copy? In channel coding, classical capacity can be additive with zero or unlimited entanglement assistance and superadditive at an intermediate budget~\cite{ZhuZhuangShor2017,ZhuZhuangHsiehShor2018}. For a fixed shared state, Wilde recently asked whether additive optimized coherent information can coexist with a superadditive trade-off at a positive communication price [Eq.~(10.1) of Ref.~\cite{Wilde2026}]. We will construct a state for which two copies suffice for a gain, even though both strategies first reach the maximum at the same cost.

\begin{figure}[t]
\centering\includegraphics[width=\columnwidth]{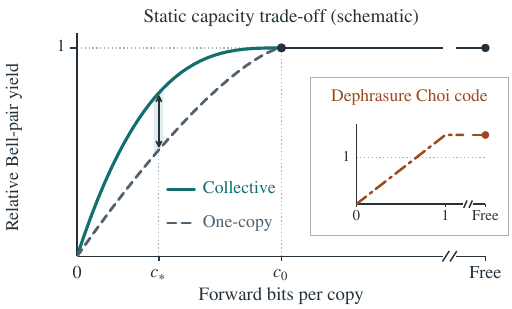}
\caption{Schematic yield in units of each state's maximum with one-copy preprocessing and free forward bits. Both yields for our state first reach $J_1$ at $c_0$. At $c_*$ the guaranteed gap is $\Delta/(2J_1)$ in plotted units, of order $h_2(\delta^4)$ at fixed overlap and price [Eq.~\eqref{eq:bridge}]. The dephrasure repetition code~\cite{LeditzkyLeungSmith2018} retains its gain on the Choi state with free forward bits (inset).}\label{fig:rate}
\end{figure}

Figure~\ref{fig:rate} compares the Bell pairs per copy attainable with at most $c$ forward bits per copy and zero net qubit rate. For $E_{\rm sep}(c)$, Alice applies an instrument with a public outcome to each copy separately. For $E_{\rm coll}(c)$, she can process several copies together. Both allow time sharing, resource conversions and asymptotic coding after preprocessing. Specifically, we will prove the following theorem.

\begin{theorem}\label{thm:static}
There is a family of degradable states $\rho_{AB}=\tfrac12\rho_N\oplus\tfrac12\rho_M$ on $\mathbb C^5\otimes\mathbb C^5$ for which $E_{\rm sep}(0)=E_{\rm coll}(0)=0$. Let $w$ denote the data label distribution and $R_N$ its reference. Both yields first reach their common maximum $J_1=D_{\to}(\rho)=I(A\rangle B)_\rho$ at the forward-bit cost $c_0$, where
\[
 J_1=\frac{H(w)-S(R_N)}{2},\qquad c_0=\frac{S(R_N)}{2}.
\]
Two-copy preprocessing already gives a strict gain at an intermediate budget. Matching sectors across a long block gives $E_{\rm coll}(c_*)-E_{\rm sep}(c_*)\ge\Delta/2>0$ at $0<c_*<c_0$, where $\Delta$ is the surplus per matched pair.
\end{theorem}

The proof requires several definitions and intermediate results and so will be given later in the main text. In this construction, both Alice and Bob can read the flag distinguishing data $N$ from a helper $M$. The equally likely sectors have purifications, in $ABR$ order,
\begin{align*}
 \ket{\psi_N}&=\sum_i\sqrt{w_i}\ket{i,i}\ket{\phi_i},\\
 \ket{\psi_M}&=\sum_{b,r}\sqrt{\pi_b\pi_r}\ket{b\oplus r,b,r}.
\end{align*}
The three data labels have priors $w=(1/4,1/4,1/2)$ and overlapping reference signals $\phi_i$. In the helper, Bob's bit $b$ and the reference bit $r$ are independent with common distribution $\pi=(1-a,a)$. Their parity flips Alice's memory with probability $p=2a(1-a)$, where $a<1/2$. Bob can reproduce the reference from his label in $N$ and by symmetry in $M$, so the state is degradable and its coherent information fixes the maximum yield~\cite{DevetakWinter2005,LeditzkyDattaSmith2018}.

Intuitively, in our construction, one copy of the state is used to supply a noisy quantum memory for the other. Alice could prepare the same memory locally by flipping a coin and erasing its record (Fig.~\ref{fig:memory}) and the helper saves forward bits because Bob already holds part of that record. 

\par\medskip\noindent\begin{minipage}{\columnwidth}
\centering\includegraphics[width=.88\columnwidth]{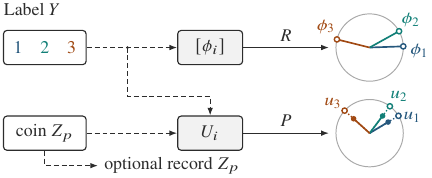}
\captionof{figure}{Each color pairs a signal with its memory. Erasing the coin flip record contracts each memory vector from the open circle to the dot. Dashed wires carry classical data.}\label{fig:memory}
\end{minipage}\par\medskip

\begin{figure*}[t]
\centering
\includegraphics[width=0.96\textwidth]{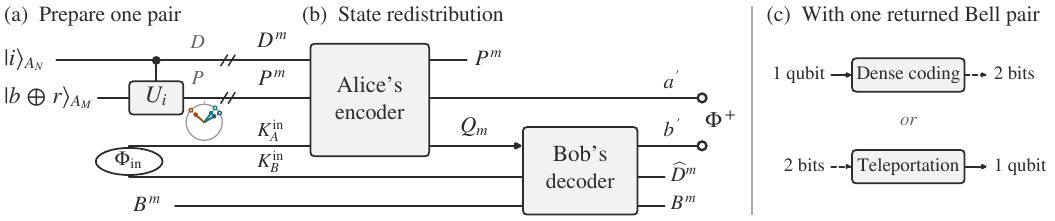}
\caption{Distillation from $N$ and $M$. (a) $U_i\ket s=\sigma_y^s\ket{u_i}$ gives $\omega_i(p)$ on $P$ after averaging over parity. (b) Alice keeps $P^m$ and sends $Q_m$ to transfer $D^m$. (c) One returned Bell pair enables dense coding~\cite{BennettWiesner1992} or teleportation~\cite{BennettEtAl1993}.}
\label{fig:circuit}
\end{figure*}

\paragraph{A mixed memory improves the information bottleneck.}
To understand the mechanism behind Theorem~\ref{thm:static}, consider the following simplified setting. Alice draws a label $Y=i$ with probability $w_i$ and prepares a signal $[\phi_i]_R$ and a memory $\omega_{i,P}$, where $[v]=\ket v\!\bra v$. Because the signals overlap, $R$ reveals only part of the label. The bottleneck rewards information about $R$ and charges for information about $Y$~\cite{TishbyPereiraBialek1999,GrimsmoStill2016,DattaHircheWinter2019}. The score in bits and its change on erasing $Z$ are~\cite[Eqs.~(8.50) and (D.124)]{Wilde2026}
\begin{align}
 j_\kappa(P)&=\kappa I(R:P)-I(Y:P),\notag\\
 j_\kappa(P)-j_\kappa(PZ)&=I(Y:Z|P)-\kappa I(R:Z|P).
 \label{eq:forgetting-identity}
\end{align}
Erasure is selective when it loses some signal information and saves more in label cost than it loses in signal reward.

How well can Alice do if she keeps a complete record? For label $i$ she can randomize over pure states $v_{ix}$ and keep the flag $x$, giving
\[
 \omega_i^{PX}=\sum_x p(x|i)[v_{ix}]_P\otimes[x]_X .
\]
Wilde's conjecture~8.10 uses these memories and asks whether their best score over all finite dimensions, $\mathsf J_\kappa^{\fp}=\sup j_\kappa(PX)$, equals the unrestricted optimum~\cite{Wilde2026}. We will prove in Theorem~\ref{thm:memory} that this conjecture is false.

For the data signals choose
\[
 \phi_{1,2}=r(\theta\pm\delta),\qquad \phi_3=r(-\theta),
\]
where $r(x)=\cos x\ket0+\sin x\ket1$ and $0<\gamma=\cos(2\theta)<1$ is the overlap at $\delta=0$. Let $P_0$ be an optimal pure-qubit memory with real rays $u_i$ (Lemma~\ref{lem:gram}). Alice flips each $u_i$ to the orthogonal ray $\sigma_y u_i$ with probability $p=\delta^4$. Keeping the coin $Z_p$ lets her undo the flip. Forgetting it contracts each memory vector along its radius. At flip probability $z$, the conditional memory and joint signal--memory state are
\begin{align*}
 \omega_i(z)&=(1-z)[u_i]+z[\sigma_y u_i],\\
 \sigma_z&=\sum_i w_i[\phi_i]\otimes\omega_i(z).
\end{align*}
With these definitions, we are now able to prove our second theorem.
\begin{theorem}\label{thm:memory}
Choose $t=1-1/\kappa>0$ below both $\gamma^2$ and $(1-\gamma)/8$. For all sufficiently small $\delta>0$, $j_\kappa(P)>\mathsf J_\kappa^{\fp}$.
\end{theorem}

\begin{figure}[!b]
\centering\includegraphics[width=\columnwidth]{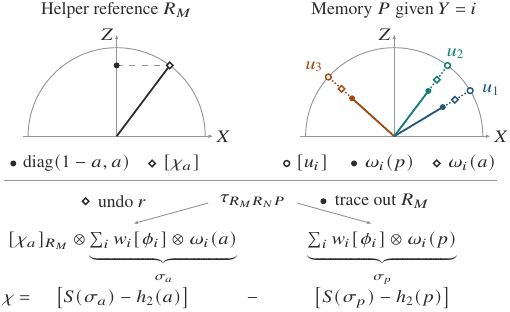}
\caption{The extra forward bits $\chi$ measure how much $R_M$ adds about $Y$ given $R_NP$. Undoing the reference flip (diamonds) leaves noise at rate $a$. Tracing out $R_M$ (dots) leaves rate $p$. Each sum pairs signals with their memories. Open circles mark unflipped rays. Displacements are enlarged.}
\label{fig:helper-cost}
\end{figure}

\begin{proof}
For a known label the coin outcomes give orthogonal memories. After averaging over labels the joint states overlap. Their entropies agree by unitary equivalence, leaving coin uncertainty
\[
 F(z)=H(Z_z|RP)=h_2(z)+S(\sigma_0)-S(\sigma_z),
\]
where $h_2$ is the binary entropy. Keeping the coin recovers $P_0$, so Eq.~\eqref{eq:forgetting-identity} gives
\begin{equation}
 [j_\kappa(P)-j_\kappa(P_0)]/\kappa=F(p)-tH(Z_p|P).
 \label{eq:gain-balance}
\end{equation}

The unflipped product rays span three dimensions of $RP$, so projecting onto the fourth with $\Pi$ certifies a flip. Lemma~\ref{lem:gram} bounds the least positive eigenvalue of $\sigma_0$ by $s_\delta\ge C\delta^2$. Each flip produces the orthogonal memory ray, so $\sigma_0+\sigma_1=\rho_R\otimes I$. Given a flip, the certification probability is $p_{\rm cert}=\Tr(\Pi\sigma_1)\le\|\rho_R\|_\infty=(1+\gamma)/2+o(1)$. The uncertified weight lies in the unflipped support. Positivity of $D(\sigma_p\Vert(1-p)\sigma_0+p\Pi)$ gives
\[
 F(p)\ge p(1-p_{\rm cert})\log_2\frac{(1-p)s_\delta}{p}
 \ge\big[(1-\gamma)/4+o(1)\big]h_2(p).
\]
At posterior $q^x$ the Gram matrix $W_x$ of the three weighted memory vectors determines both branch entropies in any dimension. For branch cost $g_\delta(W_x)$ and best qubit cost $m_{2,\delta}$ at $w$, the excess over $P_0$ is $\kappa[m_{2,\delta}-\sum_xp_xg_\delta(W_x)]$. Lemma~\ref{lem:gram} bounds this by an affine correction and $o(h_2(p))$. The linear term cancels because $\sum_xp_xq^x=w$. Using $H(Z_p|P)\le h_2(p)$ gives
\begin{align*}
 j_\kappa(P)-j_\kappa(P_0)&\ge A_{\gamma,t}\,h_2(p),\\
 \mathsf J_\kappa^{\fp}-j_\kappa(P_0)&=o(h_2(p)),
\end{align*}
for any fixed $0<A_{\gamma,t}<\kappa[(1-\gamma)/4-t]$ and sufficiently small $\delta$. Erasure also loses signal information. Revealing the mixture flag in $\sigma_p=(1-2p)\sigma_0+2p\rho_R\otimes I/2$ gives $I(R:P)\le(1-2p)I(R:P_0)$, while beating a constant memory requires $I(R:P)>0$.
\end{proof}

\paragraph{Distillation with one-copy preprocessing.}
Weight net bits, qubits, and ebits by $(\lambda,1+2\mu,1)$, with generation positive and consumption negative~\cite{HsiehWilde2010}. Write $\mathcal S_{\lambda,\mu}$ for the supremum of $\lambda C+(1+2\mu)Q+E$ over the one-copy region. At $(-c,0,E)$ and $\lambda=\mu$ this is $E-\mu c$, so
\begin{equation}
 E_{\rm sep}(c)\le\mathcal S_{\mu,\mu}(\rho)+\mu c . \label{eq:budget-line}
\end{equation}
For $\mu\ge0$ and $\mu\le\lambda\le1+\mu$, the support is the supremum of the instrument score~\cite[Eqs.~(5.16)--(5.19)]{Wilde2026}
\begin{equation}
 s=-S(D|BX)-\lambda I(X:R|B)-\mu I(D:R|BX). \label{eq:static-score}
\end{equation}
Here $X$ is public, $D$ is the instrument output, and $P$ is the environment Alice can retain. Write $\mathcal S=\mathcal S_{\lambda,\mu}$, $\kappa=1+\mu$, and $b_N=H(w)-\kappa S(R_N)$.

On the span of the data vectors $\ket{i,i}$, Wilde's normal form attains the full one-copy support. Its branch map $V_x\ket{i}=\sqrt{p(x|i)}\ket{i}_D\ket{v_{ix}}_P$ keeps the label coherently in $D$ and leaves $P$ pure given $i,x$. Conversely, every flagged pure preparation defines an instrument since $\sum_x V_x^\dagger V_x=I$. Its bottleneck memory is $XP$, giving $\mathcal S(N)=b_N+\mathsf J_\kappa^{\fp}$ throughout this price range [Theorems~8.1 and 8.3 and Eq.~(8.53) of Ref.~\cite{Wilde2026}].

How does Alice pay for a local coin in the distillation protocol? For an unflagged isometry $A_N\to DP$, purity gives $S(DB_N)=S(R_NP)$ and $I(D:R_N|B_N)=S(R_N|P)$, so Eq.~\eqref{eq:static-score} becomes
\[
 s=b_N+j_\kappa(P)-S(P|Y).
\]
Keeping the coin's purification enlarges Alice's private memory. Keeping just $P$ leaves its purifier in $D$, costing $S(P|Y)=h_2(p)$. This absorbs the gain in Eq.~\eqref{eq:gain-balance} since $F(p)\le H(Z_p|P)\le h_2(p)$.

\begin{figure}[!ht]
\centering\includegraphics[width=\columnwidth]{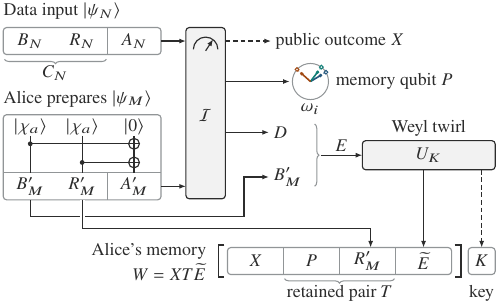}
\caption{One-input test of Fig.~\ref{fig:circuit}(a). Alice prepares the primed helper locally. The instrument on $A_NA'_M$ yields a fixed public outcome $X$ and memory $\omega_i(p)$ on $P$ given Bob's label $i$. Keeping $T=PR'_M$ lets Alice undo the reference flip (Fig.~\ref{fig:helper-cost}). Erasing the Weyl key $K$ for $DB'_M$ decouples $\widetilde E$ from $C_NXT$, so $W=XT\widetilde E$ has the score of $XT$.}
\label{fig:record-test}
\end{figure}

\begin{proof}[Proof of Theorem~\ref{thm:static}]
Choose $\mu=\kappa-1$ as in Theorem~\ref{thm:memory} and fix sufficiently small $\delta>0$ before taking the blocklength limit. On an $NM$ pair Alice flips the memory for data label $i$ using the helper parity $s=b\oplus r$. Her isometry $\ket{i,s}\mapsto\ket{i}_D\sigma_y^s\ket{u_i}_P$ lets her keep $P$ and transfer $D$ to Bob [Fig.~\ref{fig:circuit}(a)].

State redistribution consumes $q$ qubits and yields $e$ ebits per matched pair~\cite{DevetakYard2008,YardDevetak2009}. Purity gives
\begin{align}
 e+q &= H(Y|P),\notag\\
 2q &= I(D:R_NR_M|B_NB_M)=S(R_N|P)+\chi .\label{eq:redistribution}
\end{align}
The parity flip raises $H(Y|P)$ as in Fig.~\ref{fig:memory}. Teleporting the message costs $2q$ forward bits, including $\chi=I(Y:R_M|R_NP)$ for the helper reference's additional label information.

To evaluate the entropy of the label-averaged $\tau$ on $R_MR_NP$, undo $r$ on $P$ under control of $R_M$ (Fig.~\ref{fig:helper-cost}). The pure helper state $\ket{\chi_a}=\sqrt{1-a}\ket0+\sqrt a\ket1$ factors off, leaving $\sigma_a$ with Bob's flip at rate $a=p/2+O(p^2)$. Tracing out $R_M$ in the original state gives $\Tr_{R_M}\tau=\sigma_p$. At fixed label the memory rays are orthogonal and the signal pure, giving entropies $h_2(a)$ and $h_2(p)$. Thus
\begin{align}
 \chi &=[S(\sigma_a)-h_2(a)]-[S(\sigma_p)-h_2(p)]\notag\\
 &=F(p)-F(a).\label{eq:helper-charge}
\end{align}

For $m$ helpers, symmetry and weak monotonicity give $S(D|B_M^m X)=S(D|R_M^m X)\ge0$, so every helper power has zero support. The pair's surplus is $\Delta=e-(2\kappa-1)q-\mathcal S(N)-\mathcal S(M)$. With $b_*=\mathsf J_\kappa^{\fp}-j_\kappa(P_0)=o(h_2(p))$, Eqs.~\eqref{eq:gain-balance} and~\eqref{eq:helper-charge} give
\begin{align}
 \Delta &=j_\kappa(P)-\mathsf J_\kappa^{\fp}-\kappa\chi \notag\\
 &=\kappa[F(a)-tH(Z_p|P)]-b_*\notag\\
 &\ge\{\kappa[(1-\gamma)/8-t]+o(1)\}h_2(p).\label{eq:bridge}
\end{align}
At least half of the coin contribution survives the helper cost, since $F(0)=0$ and concavity give $F(a)\ge(a/p)F(p)\ge F(p)/2$. Here we used $H(Z_p|P)\le h_2(p)$ and the bound on $F(p)$. Since $\Delta\le\kappa F(a)\le\kappa h_2(a)$ and $a/p\to1/2$, the certified surplus is of order $h_2(\delta^4)$. Table~\ref{tab:tasks} gives the corresponding rates.

\begin{table}[!hb]
\caption{Rates per matched pair at equal input cost. Dense coding uses $\lambda=1+\mu$, distillation and teleportation $\lambda=\mu$. Unlisted net rates vanish.}
\label{tab:tasks}\centering
\setlength{\tabcolsep}{2.4pt}
\begin{ruledtabular}
\begin{tabular}{lccc}
Protocol & Input & Output & Gain at least\\
\hline
Redistribution & $q$ qubits & $e$ ebits & $\Delta$ ebits\\
Distillation & $2q$ bits & $(e-q)$ ebits & $\Delta$ ebits\\
Dense coding & $(q+e)$ qubits & $2e$ bits & $\Delta/\kappa$ bits\\
Teleportation & $2e$ bits & $(e-q)$ qubits & $\frac{\Delta}{2\kappa-1}$ qubits\\
\end{tabular}
\end{ruledtabular}
\end{table}

Reading the sector attains the mean support. Conversely, dephasing the purifying flag preserves $DBX$ and can only lower reference penalties. Since Bob knows the sector, the mean bounds every one-copy instrument. Fixed pairs have unlike sectors half the time. Optimizing separately on like sectors therefore gives $\mathcal S(\rho^{\otimes2})-2\mathcal S(\rho)\ge\Delta/2$ per pair for every $\mu\le\lambda\le1+\mu$.

For explicit distillation, an unlike pair yields $r=e-q$ ebits for $2q$ bits. On $NN$ use $P_0$ twice, and discard $MM$ (each occurs with probability $1/4$). Each unflipped data copy uses \mbox{$2q_0=S(R_N|P_0)$} bits for \mbox{$r_0=H(w)-S(\sigma_0)$} ebits, missing its support by $b_*$. In a long block their identical flags let both parties match almost all data and helpers in order without communication. The per-copy rates and excess above~\eqref{eq:budget-line} are
\[
\begin{array}{lccc}
 & c\ \text{(bits)} & Y\ \text{(ebits)} & \text{Excess score}\\
 \text{Fixed pairs }(c_2,Y_2) & (q_0+q)/2 & (r_0+r)/4 & (\Delta-b_*)/4\\
 \text{Long block }(c_*,Y_*) & q & r/2 & \Delta/2
\end{array}
\]
Both gains are positive by Eq.~\eqref{eq:bridge}, as is $r=H(w)+h_2(a)-S(\sigma_a)$ since $\sigma_a$ tends to a mixture of two rays. Hashing a sublinear number of data copies at rate $H(w)-S(R_N)>0$ seeds the first block at vanishing per-copy cost. Later blocks replenish the seed. Since $I(Y:P),\mathsf J_\kappa^{\fp}\ge0$, Eq.~\eqref{eq:bridge} gives \mbox{$2(c_0-c_*)=I(R_N:P)-\chi\ge\Delta/\kappa>0$}. Reaching $Y_*$ saves at least $\Delta/(2\mu)$ forward bits per copy.

Why do the yields agree at $c=0$ and $c=c_0$? Discarding helpers and teleporting the redistributed data attains $(c_0,J_1)$. Hold this state fixed as the price $\mu'\to0^+$. A memory with nonnegative bottleneck score must lose its ability to distinguish labels, since a binary readout would otherwise distinguish overlapping pure signals while leaving them undisturbed in the limit (Lemma~\ref{lem:small-price}). Removing data sites one at a time gives the all-block bound $E\le J_1+\mu'(c-c_0)+o(\mu')$, forcing $c\ge c_0$ at $E=J_1$. At a sufficiently high finite price the full label is optimal and the same converse gives $E\le\mu' c$, hence zero yield at zero budget.
\end{proof}
\paragraph{Every collective gain admits a selective-forgetting test.}
Erasing the coin flip record beats every flagged pure preparation in Theorem~\ref{thm:memory}. For independent inputs $\bigotimes_i\ket{\psi_i}_{A_iB_iR_i}$, Alice tests share $A_i$ by preparing the other purifications and applying the joint instrument to the $A$ registers (Fig.~\ref{fig:record-test}). She compares the same memory with its key retained or erased. The score $j_i(W)=\kappa I(R_i:W)-I(C_i:W)$ rewards information about $R_i$ and charges for information about $C_i=B_iR_i$, with $\kappa=1+\mu$.

\begin{theorem}\label{thm:record-deletion}
With finite-dimensional registers, suppose a joint instrument on $n\ge2$ independent inputs beats the sum of their individual optima by $\Gamma>0$ at prices $\mu\ge0$, $\mu\le\lambda\le1+\mu$. For at least two choices of $i$, Alice can test input $i$ with public outcome $X$ retained in $W$. Given the classical records, including $K$, her private memory purifies $C_i$. Erasing $K$ incurs a signal loss $I(R_i{:}K|W)>0$ and raises the score by
\begin{equation}
 j_i(W)-j_i(WK)=I(C_i{:}K|W)-\kappa I(R_i{:}K|W) \ge \frac{\Gamma}{2(n-1)} .\label{eq:record-gain}
\end{equation}
\end{theorem}

\begin{proof}
Alice keeps the environment and local purifying registers. Independence splits the signal reward by the chain rule. Comparison with the individual optima lets her exclude any input and find another where omitting registers gains at least $\Gamma/(n-1)$ and her private memory carries signal information given $X$ (End Matter).

With small independent probability Alice omits all private registers, recording the choice in $W$. This ensures positive signal loss while preserving half the gain. She implements either omission by masking the registers $E$ with a uniformly random Weyl operator~\cite{Werner2001}. Keeping $K$ recovers the full memory. Erasing it leaves $I_E/\dim E$ in tensor product with the other registers and gives exactly the omission score.
\end{proof}

As a simple example of favorable erasure, consider a Bell pair and an independent shared fair bit purified by a GHZ state. On $m$ copies, conditioning on Bob's bits and $X$ leaves him $m$ qubits, so $-S(D|BX)\le m$. Taking Alice's Bell halves as $D$ attains support $m$. Masking her input and erasing the key loses one bit about $R$ and four about $BR$, gaining $4-\kappa$ for $\kappa<4$. Thus, favorable erasure occurs even on this additive state. Whether conditioning can preserve the common maximum and the same instrument's gain while exposing a memory that beats the best eligible one-input preparation remains open.

\paragraph{Discussion.}
The dephrasure Choi state, obtained by sending half a Bell pair through the channel, retains its gain with free forward bits (Fig.~\ref{fig:rate}). Resolving the Kraus outcome leaves Alice with a purification of the channel input and can only raise mean coherent information~\cite{DevetakWinter2005}. By Pauli covariance, steering an optimizer over its uniform Pauli orbit attains the one-copy bound~\cite{HJW1993}. In the $00,11$ repetition code, one erasure contributes zero, two cost one ebit, and no erasure leaves phase-flip probability $2d(1-d)$~\cite{LeditzkyLeungSmith2018}. The channel gain has been observed experimentally~\cite{YuEtAl2020}. One parity bit and $1-2R_2$ hashing bits~\cite{BDSW1996,DevetakWinter2005} per pair suffice. Thus one forward bit per copy suffices for a yield $R_2$ above the separate maximum (End Matter).

Noisy preprocessing also improves one-way secret key distillation~\cite{KrausGisinRenner2005,RenesSmith2007,SmithRenesSmolin2008}. A mixed private ensemble for the platypus channel assigns weight $1/2$ to its one-dimensional input sector~\cite{Leditzky2023Platypus}, compared with $1/3$ in the maximally mixed channel-input marginal used in the Choi construction, and an assisting erasure channel can raise its coherent information~\cite{LeditzkyEtAl2023}. The switch-channel encoder of Ref.~\cite{ZhuZhuangShor2017} [Eq.~(1)] can choose the same branch on every use. Its Choi state supplies each branch only half the time, so distillation requires a separate optimization.

In our state the helper yields no Bell pairs on its own, yet its correlations reduce what Alice must tell Bob at an intermediate yield. Separate processing already reaches the maximum at the least possible cost.

\begin{acknowledgments}
AMP thanks CMR and BRP for helpful conversations, and Graeme Smith for pointing out connections with noisy private ensembles and the platypus channel.
\end{acknowledgments}
\paragraph{AI use disclosure.} ChatGPT (OpenAI) assisted with proof development and checking, manuscript drafting, equation typesetting, and figure generation.
\clearpage
\bibliography{references}
\clearpage
\section*{End Matter}
\setlength{\abovedisplayskip}{6pt plus 2pt minus 3pt}
\setlength{\belowdisplayskip}{6pt plus 2pt minus 3pt}
\setlength{\abovedisplayshortskip}{6pt plus 2pt minus 3pt}
\setlength{\belowdisplayshortskip}{6pt plus 2pt minus 3pt}
\paragraph{Flagged preparations in Theorem~\ref{thm:memory}.}
For each flag with $p_x>0$ set $q_i^x=w_i p(x|i)/p_x$. The memory and joint Gram matrices are~\cite{JozsaSchlienz2000}
\[
 (W_x)_{ij}=\sqrt{q_i^x q_j^x}\langle v_{ix}|v_{jx}\rangle,\qquad T(W)_{ij}=\langle\phi_i|\phi_j\rangle W_{ij}.
\]
Their spectra give both entropies. Every $W\ge0$ with diagonal $q$ factors into weighted rays, including boundary priors. Since $q_1+q_2$ has fixed mean, the branch cost is
\[
 g_\delta(W)=S(T(W))-tS(W)-(1-t)[H(q)-q_1-q_2].
\]
Write $m_{2,\delta}$ for its minimum over matrices of rank at most two with diagonal $w$. 

\newtheorem{gramlemma}{Lemma}
\begin{gramlemma}[Supporting plane]\label{lem:gram}
Fix $0<t<\min\{\gamma^2,1/8\}$. For sufficiently small $\delta>0$ there exist $\beta_\delta$ and $0\le r_\delta=o(h_2(p))$ such that every $3\times3$ density matrix $W$ satisfies
\begin{equation}
 g_\delta(W)\ge m_{2,\delta} +\beta_\delta\!\cdot\!(\diag W-w)-r_\delta . \label{eq:supporting-plane}
\end{equation}
A minimizing qubit preparation at $w$ has real rays and a rank-three joint state with least positive eigenvalue at least $C\delta^2$ for a constant $C>0$.
\end{gramlemma}

\paragraph{Proof of Lemma~\ref{lem:gram}.}
\emph{The unique minimum at $\delta=0$.} For a binary prior $(s,1-s)$ and memory overlap $d$, put $m=(2s-1)^2$, $x=m+(1-m)d^2$, and $b=(1-\gamma^2)m+\gamma^2x$. The Gram spectra $(1\pm\sqrt x)/2$ and $(1\pm\sqrt b)/2$ give, for $H(y)=h_2((1+\sqrt y)/2)$,
\[
 f(m,x)=H(b)-tH(x)-(1-t)H(m).
\]
Write $A(y)=-2\ln(2)H'(y)=\int_0^1(1-yv^2)^{-1}dv$. Cauchy--Schwarz gives $AA''>2(A')^2$, so $1/A$ is strictly concave. At $x=m<1$, $f=0$ and $f_x<0$, whereas $f_x\to+\infty$ as $x\to1$. At an interior minimum $t=\gamma^2 A(b)/A(x)$, and concavity of $1/A$ at $b$ gives
\[
 2\ln(2)f_m=(1-t)A(m)-(1-\gamma^2)A(b)>0.
\]
The minimum $f_*$ occurs at $m=0$ ($m=1$ gives zero). There $\partial_d f$ has the sign of $t-Q(d)$ for $Q(d)=\gamma\operatorname{atanh}(\gamma d)/\operatorname{atanh}d$. The ratio of numerator and denominator derivatives decreases strictly, so $Q$ decreases with $Q'<0$ from $\gamma^2$ to zero. At the unique solution $Q(d_0)=t$ both curvatures are positive ($f_{ss}=8f_m$), and symmetry gives $f_{sd}=0$.

Write $q=(s\alpha,s(1-\alpha),1-s)$ and let $A_2$ be the leading $2\times2$ block of $W$. At $\delta=0$ the first two signals coincide. Mix the binary pairs $(1,3)$ and $(2,3)$ at prior $(s,1-s)$ with weights $\alpha,1-\alpha$. Their memory marginals differ in trace norm by $2s\sqrt{1-|\langle u_1|u_2\rangle|^2}$. Concavity of $S(R|P)$ and Pinsker's bound for $(1-t)S(P)$ give
\[
 g_0(W)\ge \alpha f_{13}+(1-\alpha)f_{23} +(1-t)\left[s(1-h_2(\alpha))+\frac{2\det A_2}{\ln2}\right].
\]
Attaining $f_*$ forces $s=\alpha=1/2$ and $\det A_2=0$. The first two rays therefore coincide and have overlap $d_0$ with the third, determining $W_0$ up to ket phases.

\emph{Reduction to real qubit rays.} The squared overlaps $a_{ij}=|\langle u_i|u_j\rangle|^2$ determine the Bloch Gram matrix with entries $2a_{ij}-1$. Positive semidefiniteness describes the whole compact convex feasible set, since any three unit vectors in $\mathbb R^3$ are Bloch vectors. At fixed $q$, $\Tr[T(W)^2]$ and $\det T(W)$ are affine in the $a_{ij}$ by the real signal overlaps and $2\operatorname{Re}\prod_{\mathrm{cyc}}\langle u_i|u_j\rangle=\sum_{i<j}a_{ij}-1$. They fix the trace-one joint spectrum. Two affine constraints in three variables leave a line from any interior point to the boundary along which $\Tr W^2$, also affine, weakly decreases and qubit entropy weakly increases. The boundary Gram matrix is singular, so rotating its Bloch plane to the $xz$ plane gives real kets of no greater cost.

\emph{Uniform curvature at fixed $\gamma,t$.} Set $u_3=\ket0$, $u_{1,2}=r(y\pm z)$, $y_0=\arccos d_0$, and $v=(q_1,q_2,y)$. The simple zero eigenvalue of $T(W_0)$ continues analytically as $\ell$. Put $h=(\delta,z)$, $\phi=r(\theta)$, $u=r(y)$. At $h=0$ the weighted product columns annihilate $(\sqrt{1-\alpha},-\sqrt\alpha,0)$. Their first variation on this vector is $2\sqrt{s\alpha(1-\alpha)}(\delta\phi'\otimes u+z\phi\otimes u')$. Adjusting the null vector cancels the projection into the column span of $\phi\otimes u$ and $\phi_3\otimes u_3$. The squared residual gives $\ell=Q_v+O(\|h\|^3)$, where for $\mathbf k=(\sin(2\theta)\cos y,\gamma\sin y)$ and $G=1-\gamma^2\cos^2y$,
\[
 Q_v(h)=4s\alpha(1-\alpha)\left(\|h\|^2-\frac{(\mathbf k\cdot h)^2}{G}\right).
\]
Since $G-\|\mathbf k\|^2=(1-\gamma^2)\sin^2y>0$ near $y_0$, $\ell$ lies between positive multiples of $\|h\|^2$ on a small closed ball about $(w_1,w_2,y_0)$. Fixing $\gamma,t$ fixes the ball and bounds. Taylor expansion in $h$ with coefficients analytic in $v$ gives uniformly
\[
 \ell_v,\ell_{vv}=O(\ell),\quad \ell_z,\ell_{vz}=O(\sqrt\ell),\quad \ell_{zz}\ge b_0>0 .
\]
For the singular entropy term the chain rule gives
\[
 (\ln2)D^2(-\ell\log_2\ell)=-(1+\ln\ell)D^2\ell-(D\ell)(D\ell)^{\mathsf T}/\ell .
\]
Its $vv$ and $vz$ blocks vanish uniformly as $h\to0$, while its $zz$ entry diverges since $\ell_z^2/\ell$ stays bounded.

The remaining cost is smooth and at $h=0$ equals the binary cost plus $(1-t)s[1-h_2(\alpha)]$, with positive Hessian in $v$ at the minimum. On a smaller convex chart the full Hessian $\mathcal H$ has $\mathcal H_{vv}\ge 2cI$ and $\|\mathcal H_{vz}\|\le M$ for some $c>0$ and finite $M$. For small $h\ne0$, $\mathcal H_{zz}\ge c+M^2/c$, so the Schur complement gives $\mathcal H\ge cI$~\cite{BoydVandenberghe2004}.

\emph{Fixed-prior minimization.} Set $\varepsilon=\delta\log_2(1/\delta)$. The $O(\delta)$ trace-norm change in $T(W)$ gives $\|g_\delta-g_0\|_\infty=O(\varepsilon)$ by entropy continuity~\cite{Audenaert2007}. Outside a small neighborhood of the minimizing phase orbit, compactness gives $g_0-f_*$ a positive lower bound. Thus a real qubit minimizer $W_*$ at $q=w$ lies in the chart for small $\delta$. Applying binary curvature at $d_0$ to the overlaps $\cos(y\pm z)$ gives $g_0-f_*\ge c[(y-y_0)^2+z^2]$ since $\sin y_0\ne0$. Comparison with $W_0$ bounds this difference by $O(\varepsilon)$, hence the displacement by $O(\sqrt\varepsilon)$.

\emph{A supporting plane for all qubit priors.} Take $\beta_\delta$ to be the two prior derivatives at $W_*$ at fixed rays, extended by zero. Smoothness and $\ell_v\log\ell=O(\varepsilon|\log\varepsilon|)$ give $\|\beta_\delta\|=O(\sqrt\varepsilon)$. Subtracting this slope makes $W_*$ stationary in all four coordinates. Since $\|q-w\|_1\le2$ on the whole simplex, the tilted cost differs uniformly from $g_0$ by $O(\sqrt\varepsilon)$, less than half the exterior gap for small $\delta$ at fixed $\gamma,t$. Its global minimum over all qubit priors thus has a real representative in the chart, where strong convexity gives
\[
 \widehat g(W):=g_\delta(W)-\beta_\delta\cdot(q-w)\ge m_{2,\delta}\qquad(\operatorname{rank}W\le2).
\]

\emph{The error from a third memory dimension.} At an unrestricted minimum of $\widehat g$, $g_0(W)-f_*=O(\sqrt\varepsilon)$ and $q\to w$. The mixing bound and $\Tr A_2\to1/2$ give $\lambda_{\min}(A_2)=O(\det A_2)=O(\sqrt\varepsilon)$. For $\lambda_3=\lambda_{\min}(W)>0$, pad a unit least eigenvector $v_0$ of $A_2$ with zero. Put $D=[v_0]-W$, $d=\diag D$, $e=(1,1,0)$. With base-two logs, stationarity gives
\begin{align*}
 0=\partial_D\widehat g(W)={}&t\Tr[D\log W]-\Tr[T(D)\log T(W)]\\
 &+d\cdot[(1-t)(\log q+e)-\beta_\delta].
\end{align*}
The diagonal term is bounded since $q\to w>0$, $\beta_\delta\to0$, and $\|d\|_1\le2$. Positivity and unitality give $T(W)\ge\lambda_3 I$. Expanding $D$ leaves entropies at most $\log_2 3$. With $\zeta=T([v_0])$, scalar Jensen in the spectral measure of $T(W)$ gives
\begin{align*}
 t\log\lambda_3 &\le t\langle v_0|\log W|v_0\rangle=\Tr[\zeta\log T(W)]+O(1)\\
 &\le\log\Tr[\zeta T(W)]+O(1).
\end{align*}
Self-adjointness of the real Schur multiplier gives $\Tr[\zeta T(W)]=\langle v_0|T^2(W)|v_0\rangle$. Since $T^2$ fixes the diagonal and multiplies entry $12$ by $\cos^2(2\delta)$, this is $\lambda_{\min}(A_2)+O(\delta^2)=O(\sqrt\varepsilon)$. Thus $\lambda_3=O_t(\varepsilon^\nu)$ for $\nu=1/(2t)$. The renormalized rank-two truncation has trace distance $\lambda_3$. Both $T$ and $\diag$ contract it, so entropy continuity gives $r_\delta=O_t(h_2(\lambda_3)+\lambda_3)$. Since $\nu>4$,
\[
 \frac{r_\delta}{h_2(\delta^4)}
 =O_t\!\left(\delta^{\nu-4}[\log(1/\delta)]^\nu\right)\to0 .
\]

\paragraph{Yields at $c=0$ and $c=c_0$ in Theorem~\ref{thm:static}.}
Fix the state, including $\delta>0$. Set $V=XP$, $\xi=1+\mu$, $\eta=\lambda-\mu$, and $b=S(A)-\xi I(A:R)$. With the Stinespring environment in $P$, $DPBR$ is pure given $X$, so
\begin{align}
 s={}&b+\xi I(R:V)-S(P|X)\notag\\
 &-(1-\eta)I(B:X)-\eta I(BR:X).\label{eq:named-shift}
\end{align}
Pair $N$ with an independent partner $L$. Supplying $\ket y$ at Alice's data port gives partner score $s_L^y$ and marginal $\tau^y_{R_LV}$. Reading $Y=y$ from $B_N$ in the joint trial gives $[\phi_y]_R\otimes\tau^y_{R_LV}$, with $R=R_N$. Averaging recovers the original $RR_LV$ state. The private charge rises by $S(P|X)-S(P|XY)=I(Y:P|X)$. For $G=B,BR$, reading $B_N$ gives $I(G_NG_L:X)\ge I(Y:X)+I(G_L:X|Y)$. The public weights sum to one, so the total charge rises by at least $I(Y:V)$. Since $R_L$ is independent of $YR$ and $R$ factors given $Y$,
\[
 I(RR_L:V)-I(R_L:V|Y)=I(R:V)-I(Y:R_L|RV).
\]
Dropping the last term and using $b=b_N+b_L$ gives $s\le b_N+j_\xi(V)+\sum_y w_y s_L^y$. Let $\mathcal B(\xi)=\sup_V j_\xi(V)$ over all finite-dimensional memories prepared from $Y$. Iteration bounds a word with $k$ data sites by $k[b_N+\mathcal B(\xi)]$ since helpers have zero support. Dephasing the reference flags preserves $DBX$ and can only lower the penalties. Bob knows the flags, so averaging with $\mathbb E k=n/2$ bounds every block's support per copy by $[b_N+\mathcal B(\xi)]/2$.

\begin{gramlemma}[Small-price bound]\label{lem:small-price}
For the fixed data ensemble, $\mathcal B(1+\mu)=o(\mu)$ as $\mu\to0^+$.
\end{gramlemma}

A constant memory scores zero. For nonnegative scores, conditional independence gives $I(Y:V|R_N)\le\mu I(R_N:V)\le\mu H(Y)$. If two memories stayed separated in trace distance as $\mu\to0^+$, binary Helstrom measurements would separate their bit laws. Along a subsequence all three laws converge. By data processing and continuity the limiting bit $T$ satisfies $I(Y:T|R_N)=0$. Exact recovery maps each $[\phi_i]$ to $[\phi_i]\otimes p(T|i)$~\cite{HaydenEtAl2004}. Fidelity monotonicity forces all bit laws to agree since the signals overlap, contradicting separation. Thus $\rho_{YV}$ approaches $\rho_Y\otimes\rho_V$ in trace distance. Continuity of $S(Y|V)$ at fixed $|Y|$ gives $I(Y:V)=o(1)$ uniformly in memory dimension~\cite[Lemma~2]{Winter2016}. Since $j_{1+\mu}(V)\le\mu I(Y:V)$, the lemma follows. Here $b_N/2=J_1-\mu c_0$.

For zero budget, prepare two signals independently given $Y$. The three rays $\phi_y^{\otimes2}$ are linearly independent (Vandermonde), so scaled reciprocal vectors read $Y$ with common probability $\epsilon>0$ and otherwise give an erasure~\cite{Chefles1998,Chefles2001}. The chain rule and data processing give $2I(Y:R_N|V)\ge I(Y:R_1R_2|V)\ge\epsilon H(Y|V)$. At $\xi\ge2/\epsilon$, $j_\xi(Y)-j_\xi(V)=\xi I(Y:R_N|V)-H(Y|V)\ge0$, hence $b_N+\mathcal B(\xi)=0$.

\paragraph{Selecting an input in Theorem~\ref{thm:record-deletion}.}
Given $X$, the private part of $U_i=XPDC_{-i}$ purifies $C_i$. Order any $m$ first, then $L=\{i\ne m:I(R_i:U_i|X)>0\}$, then the rest. Set $W_i=VR_{<i}$ in this order, $F_i=\mathcal S_{\lambda,\mu}(\rho_i)-b_i$, and $c_i=(1-\eta)I(B_i:X)+\eta I(C_i:X)$. In the local trial on $m$, Alice keeps $PC_L$ and transfers the other purifiers. Equation~\eqref{eq:named-shift} gives
\[
 s_m-b_m=\kappa I(R_m:VC_L)-S(PC_L|X)-c_m\le F_m .
\]
Independence splits the joint reward and bounds the public charge below by $\sum_i c_i$. Subtracting the local score cancels $S(P|X)$. The extra term is $S(C_L|V)-\kappa I(R_m:C_L|V)\le S(C_L|VR_m)\le\sum_{i\in L}S(C_i|W_i)$ by strong subadditivity~\cite{LiebRuskai1973}. Hence
\[
 \Gamma+\sum_{i\ne m}F_i \le\sum_{i\ne m}[\kappa I(R_i:W_i)-c_i]+\sum_{i\in L}S(C_i|W_i).
\]

Outside $L$, $X$ alone attains the summand. Inside $L$ it is $j_i(W_i)+h_i-c_i$ for $h_i=S(C_i)$. Conditional purity gives
\[
 j_i(U_i)+h_i-c_i =\kappa I(R_i:A_i)-h_i-\lambda I(R_i:X|B_i)\le F_i.
\]
Taking $P=A_i$ privately with trivial $D$ and $X$ attains the first two terms since $S(A_i)=h_i$. Omitting $E_i=DB_{-i}R_{>i}$ therefore gains $\sum_{i\in L}[j_i(W_i)-j_i(U_i)]\ge\Gamma$ in total, hence at least $\Gamma/(n-1)$ at one site.

\paragraph{Dephrasure Choi state in Fig.~\ref{fig:rate}.}
At phase-flip rate $d=1/9$ and erasure rate $q=1/3$, a diagonal input attains the separate optimum $\max_{0\le s\le1/2} f(s)$~\cite[Eqs.~(8)--(10)]{LeditzkyLeungSmith2018}, where $v\le1/2$ and
\[
 f(s)=\tfrac13[h_2(s)-2h_2(v)],\quad v(1-v)=4d(1-d)s(1-s).
\]
Both entropies increase with $s$. Bounding the first at right and the second at left endpoints of a $2^{-21}$ mesh gives $0.004582<f(0.07)\le\max f<0.004583$. Weighting the repetition code's erasure outcomes gives the time-shared rate $R_2\min(c,1)$ with
\[
 R_2=\tfrac12\big\{(1-q)^2[1-h_2(2d(1-d))]-q^2\big\}\approx0.007342 .
\]
\end{document}